\documentclass[11pt,a4paper]{article}
\pdfoutput=1

\usepackage{jcappub}

\usepackage{graphicx}
\usepackage{enumitem}
\usepackage{amsmath}

\usepackage{booktabs}
\usepackage{array}
\usepackage{color}
\usepackage{amssymb}
\usepackage{mathrsfs}
\usepackage{MnSymbol}

\usepackage[T1]{fontenc}
\usepackage[latin1]{inputenc}
\usepackage[table]{xcolor}  
\usepackage{verbatim}
\usepackage{caption}
\usepackage{subcaption}
\usepackage{rotating}
\usepackage{dcolumn}
\usepackage{bm}

\newcommand{\ipleft}{\langle\kern-0.2em\langle}
\newcommand{\ipright}{\rangle\kern-0.2em\rangle}

\newcommand{\bx}{{\mathbf{x}} }
\newcommand{\bn}{{\mathbf{n}} }
\newcommand{\bk}{{\mathbf{k}} }
\newcommand{\GG}{{\mathcal{G}} }

\newcommand{\vba}{{\overline{v}} }

\renewcommand{\leq}{\leqslant}
\renewcommand{\geq}{\geqslant}

\newcommand{\para}[1]{\par\vspace{2mm}\noindent\textit{{#1}---}}

\newcolumntype{s}{>{$\displaystyle}l<{$}}
\newcolumntype{t}{>{$\displaystyle}c<{$}}
\newcolumntype{u}{>{$\displaystyle}r<{$}}
\newcolumntype{v}{>{$\displaystyle}m{4cm}<{$}}

\newcolumntype{d}{D{!}{\;\pm\;}{-1}}

\newcommand{\ben}{\begin{equation}}
\newcommand{\een}{\end{equation}}
\newcommand{\bea}{\begin{eqnarray}}
\newcommand{\eea}{\end{eqnarray}}
\newcommand{\ba}{\begin{array}}
\newcommand{\ea}{\end{array}}
\newcommand{\bit}{\begin{itemize}}
\newcommand{\eit}{\end{itemize}}

\newcommand{\bXs}{\mathbf{X}}

\newcommand{\cmbact}{CMBACT}

\def\setsize{\csname @setfontsize\endcsname \setsize}

\title{The bispectrum of cosmic string temperature fluctuations including
recombination effects}
\author[1]{Donough Regan}
\author[1,2]{and Mark Hindmarsh}
\affiliation[1]{Astronomy Centre, University of Sussex,
Falmer, Brighton, BN1 9QH, UK}
\affiliation[2]{Department of Physics and Helsinki Institute of Physics, P.O.\ Box 64, 00014 Helsinki University, Finland} 

\emailAdd{d.regan@sussex.ac.uk}
\emailAdd{m.b.hindmarsh@sussex.ac.uk}
\abstract{We calculate the cosmic microwave background temperature bispectrum from cosmic strings, for the first time including the contributions from the last scattering surface, 
using a well-established Gaussian model for the string energy-momentum correlation functions, 
and a simplified model for the cosmic fluid. 
We check our approximation for the integrated Sachs-Wolfe (ISW) contribution against the bispectrum obtained from the full sky map of the cosmic string ISW signal used by the {\it Planck} team, obtaining good agreement. 
We validate our model for the last scattering surface contribution by comparing the predicted temperature power spectrum with that obtained from a full Boltzmann code treatment 
applied to the Unconnected Segment Model of a string network. 
We find that including the last scattering contribution has only a small impact on the upper limit on the string tension resulting from the bispectrum at Planck resolutions, and argue that the bispectrum is unlikely to be competitive with the power spectrum at any resolution.
}

\begin{document}	
\maketitle

\section{Introduction}
Cosmic strings are linear topological defects which may be formed at symmetry-breaking phase transitions in the early Universe \cite{Kibble:1976sj}. 
Observation of the power spectrum of the cosmic microwave background (CMB) radiation 
has gradually revealed that cosmic strings cannot be the dominant source of the seeding of large-scale structure in the Universe. 
All-sky CMB satellite experiments such as WMAP and {\it Planck} have shown that the cosmic strings contribute less than $3\%$ of the power observed \cite{Hinshaw:2012aka,Ade:2013xla}.  
Nevertheless, interest in strings remains very well motivated, as cosmic strings are a by-product of many particle physics models with extra gauge symmetries \cite{Jeannerot:2003qv}, and cosmic strings are generically formed at the end of brane inflation \cite{Sarangi:2002yt}. A detection of cosmic strings would provide important clues about the theory of the very early universe and, perhaps, as to the validity of the string theory paradigm.

There are numerous methods by which the existence of cosmic strings may be tested. 
(for a review see, for example, \cite{2009NuPhS.192...68S,1995RPPh...58..477H,Hindmarsh:2011qj}). 
However, constraints from many of these methods are either weaker than those from the CMB power spectrum, or subject to uncertainties in our understanding of cosmic string evolution.
In this paper we assess the widely-held belief that the non-gaussianity of the cosmic string CMB signal can provide stronger constraints than the power spectrum. For example, in \cite{Amsel:2007ki,Danos:2008fq}, an edge detection (Canny) algorithm was suggested as a method to search for cosmic strings in CMB anisotropy maps, suggesting that such a probe may prove superior to existing measures by up to an order of magnitude.

This program has been undertaken by the authors in previous papers, specifically \cite{Hindmarsh:1993pu,Hindmarsh:2009qk,Hindmarsh:2009es,Regan:2009hv} in the context of the integrated Sachs-Wolf (ISW) signature of the CMB, and in \cite{2015JCAP...03..008R} for the case of matter perturbations. The latter paper established that matter polyspectra are unlikely to provide competitive constraints to those obtained using the CMB, while the former papers established that the CMB ISW bispectrum is suppressed, ultimately by the approximate time reversal invariance of the string network.\footnote{B. Wandelt, private communication.} 


This symmetry suppression ensures that the bispectrum of the ISW signature will not be competitive with power spectrum constraints. These analytic considerations were further established by simulations of the post-recombination signal using stacked maps \cite{Ade:2013xla}. The four point correlation function (trispectrum) of the ISW effect was also considered in \cite{2010arXiv1012.6039F}, and shown not to be symmetry-suppressed. However,  the relatively low signal to noise and difficulty of measurement means that the trispectrum is also a sub-optimal measure. 

The question remains whether the inclusion of effects from the recombination surface may alter these conclusions
- particularly in the case of the bispectrum. It is known for the power spectrum that recombination effects dominates over the ISW signal for angular scales in the range $200\lesssim \ell \lesssim 500$. The recombination surface CMB fluctuations are generated by fluid perturbations, and the simple arguments leading to the suppression of the ISW bispectrum do not apply. Hence it is important to quantify the contribution. In addition the most recent constraints obtained using {\it Planck} CMB data ignored this effect. Therefore, a sizeable recombination surface contribution may change constraints dramatically.

Cosmic string constraints are generally characterised in terms of the dimensionless mass per unit length, $G\mu$, where G is Newton's constant, and $\mu$ represents the mass per unit length, which is proportional to the square of the symmetry breaking scale. Current power spectrum constraints on cosmic strings result in the 95\% confidence bound $G\mu \lesssim (1.5-3.2)\times10^{-7}$ (with the range due to uncertainties in modelling between Abelian-Higgs and Nambu-Goto simulations), with the corresponding ISW bispectrum constraint giving $G\mu\lesssim 8.8\times 10^{-7}$.

The contribution to the bispectrum from the last scattering surface has thus far been largely ignored due to its computational complexity. While numerical approaches to compute the power spectrum may avail of orthogonalisation techniques (such as the Cholesky decomposition) to decompose the unequal time correlators (UETCs) source term into a sum of coherent sources, this is not feasible for the bispectrum, for which 3-point correlation functions of the string energy-momentum tensor would need to be computed. 
In order to make progress, we make a number of simplifying assumptions. We model the 3-point UETCs using a Gaussian approximation for the string network. 
We ignore the baryon content of the cosmic fluid and treat the Silk damping using a damping envelope rather than computing with a full Boltzmann code. Compensation effects induced by the cosmic strings on super horizon scales are fitted using a fitting function determined in \cite{1992PhRvL..68.2121A} and discussed in some detail in \cite{Veeraraghavan:1990yd,2015JCAP...03..008R}. 

With our approximations, we may find the fluid perturbations with a Green's function method applied to the source functions (the model UETCs). 
The Green's functions are calculated numerically and the resulting (last-scattering surface) bispectrum may then be readily computed in an efficient manner. In order to obtain the CMB bispectrum from this bispectrum a weighted integral over spherical bessel functions must be performed. This integration is practically intractable unless the bispectrum is separable, i.e. unless the bispectrum can be written as a sum of terms of the form $f(k_1)g(k_2)h(k_3)$. Therefore, we employ a modal separation technique \cite{2010PhRvD..82b3502F,2010PhRvD..82b3520R} which allows one to take a grid of bispectrum values and write it in terms of a sum of one dimensional partial waves.
In a further approximation, we neglect correlations between the ISW signal and that from the last scattering surface, and add them to obtain an estimate of the full CMB power spectrum and bispectrum. 
In order to test the approximations, a comparison is made between the power spectrum and that computed using the full Boltzmann code using \cmbact  \cite{Pogosian:1999np}. 

Our computation of the ISW signature follows the approach outlined in \cite{Hindmarsh:2009qk,Regan:2009hv}, and is carried out using a flat-sky approximation, with the full sky CMB power spectrum and bispectrum recovered using the Limber approximation.  We expect that the flat-sky approximation gives reasonable results at angular scales of order 1 degree or less. As with the bispectrum due to last scattering surface effects, the ISW bispectrum is again decomposed into a sum over partial waves, allowing the calculation of the signal to noise to be readily computed.

Combination of the bispectra from recombination to that due solely to the ISW effect are shown to result in only a moderate decrease of the $1\sigma$ error bar from $\Delta (G\mu/10^{-6})^3 \approx 0.15$ to  $\Delta (G\mu/10^{-6})^3 \approx 0.11$, confirming that indications from the ISW only calculation that the bispectrum is a sub-optimal probe (compared to the power spectrum) remain true when all contributions are taken into account.

The paper is outlined as follows: In Section \ref{sec:method} we describe the methods used to compute the temperature fluctuations due to cosmic strings. We firstly describe the Gaussian model for the computation of the unequal time correlators. We then recapitulate the calculation of the ISW signal, and outline the method used in this paper to compute the contributions from the last scattering surface. In Section \ref{sec:power} we describe our computation of the power spectrum, describing the ISW spectrum first, followed by the contributions from recombination. We compare the total power spectrum to that obtained using a full Boltzmann code in order to test the validity of the approximations used. Section \ref{sec:bisp} is dedicated to the calculation of the bispectrum from both the ISW and recombination contributions, while in Section \ref{sec:results} we present and discuss the results of these computations. Finally in Section \ref{sec:conclusions} we present our conclusions. In addition we include in Appendix \ref{sec:AppA} a discussion of the separable basis technique used to project the recombination bispectra to their CMB counterparts.

\section{Temperature Fluctuations }\label{sec:method}
In the synchronous gauge, given by the metric $ds^2=a^2(\eta)[-d\eta^2 +(\delta_{ij}+h_{ij})dx^i dx^j]$, using the thin scattering surface approximation (whereby baryons and radiation act as a tightly coupled fluid until recombination which occurs instantaneously), the temperature anisotropy in direction $\hat{\bf n}$ on the sky is given by \cite{1997PhRvL..79.2180S} 
\begin{align}\label{eq:tempflucs}
\frac{\Delta T}{T}({\hat{\bf n}})=\frac{\delta_r}{4}(\eta_{\rm dec}) + \hat{\bf n}\cdot{\bf v}_r (\eta_{\rm dec})- \frac{1}{2}\int_{\eta_{\rm dec}}^{\eta_0}d\eta \dot{h}_{i j}(\eta)\hat{n}^i \hat{n}^j \,,
\end{align}
where $\eta_{\rm dec}$ and $\eta_0$ represent the conformal times at recombination and today, respectively. 
The first two terms represent the intrinsic and doppler terms at the last scattering surface respectively, with $\delta_r$ and ${\bf v}_r$ denoting the energy density ond velocity of the photon-baryon fluid. The third term, dubbed the integrated Sachs Wolfe (ISW) effect, describes the integral of the metric fluctuation along the line of sight ($\bx(\eta)=\bx(\eta_{\rm dec}) +\hat{\bf n} (\eta-\eta_{\rm dec} )$), with the overdot denoting differentiation with respect to conformal time, $\eta$. 
The sources of the fluctuations are described by their stress-energy tensor, $\Theta_{\mu\nu}$, given, in the case of a cosmic string with spacetime trajectory $X^\mu=(\eta,{\bf X}(\sigma,\eta))$ (where $(\sigma,\eta)$ denote the world-sheet coordinates), by
\begin{align}\label{eq:stressenergy}
\Theta_{\mu \nu}(\bx,\eta)=\mu \int d\sigma \left( \epsilon \dot{X}^{\mu} \dot{X}^{\nu}-\epsilon^{-1} {X'}^{\mu} {X'}^{\nu}\right) \delta^{(3)} (\mathbf{x-X})\,, 
\end{align}
where prime denotes derivatives with respect to $\sigma$, $\epsilon=\sqrt{ {\mathbf{X'}^2}/{( 1-{\dot{\mathbf{X}}^2}) }  } $ and we impose $\dot{\mathbf{X}}\cdot\mathbf{X'}=0$. 

\subsection{Gaussian Model}\label{eq:gaussmodel}
Computation of the power spectrum and bispectrum of the temperature fluctuations necessitates the calculation of the unequal time correlators of the energy-momentum tensor. In the case of the power spectrum, numerical codes such as CMBACT are based on the unconnected segment model (USM), where the string network is modelled as a set of straight segments which are randomly positioned and oriented, and which move with fixed velocity at each conformal time, $\vba(\eta)$, at right angles to the orientation. Despite recent analytical advances, \cite{Avgoustidis:2012gb}, in the case of the power spectrum, the calculation of unequal time three point correlators in the case of the bispectrum is prohibitive with this approach. Instead we use a Gaussian model, first described in \cite{Hindmarsh:1993pu}, where the string network is described - for modes well inside the horizon - as an ensemble of randomly placed strings with a Gaussian distribution for the random fields $\dot\bXs$, $\bXs'$. 

We denote the relevant correlation functions as:
\bea\label{eq:correls}
\Gamma(\sigma_-,\eta) &=& \langle [\bXs(\sigma,\eta)- \bXs(\sigma',\eta)]^2\rangle\,, \qquad
\Pi(\sigma_-,\eta) = \langle (\bXs(\sigma,\eta)- \bXs(\sigma',\eta))\cdot\dot{\bXs}(\sigma',\eta)\rangle\,, \nonumber\\
V(\sigma_-,\eta) &=& \langle  \dot{\bXs}(\sigma,\eta)  \cdot\dot{\bXs}(\sigma',\eta) \rangle\,,
\eea
where we write $\sigma_-\equiv \sigma - \sigma'$\,. The asymptotic small scale limit for these functions (dropping the explicit reference to the time dependence) is given by
\ben\label{eq:small_scale}
\Gamma(\sigma)\approx \overline{t}^2\sigma^2,\,\Pi(\sigma)\approx c_0 \sigma/(2\xi),\,V(\sigma)\approx \overline{v}^2,
\een 
with $\xi$ denoting the correlation length, which in the scaling regime (satisfied in the deep radiation and deep matter eras) obeys the relation $\xi\propto \eta$. We shall, therefore, often express the correlation length in the form $\xi(\eta)=\alpha(\eta) \eta$.
The constant $c_0$ is small due to approximate time-reversal symmetry, and may be expressed in the form \cite{Regan:2009hv,Martins:2000cs} 
\bea
c_0=\xi \langle \dot{X}^i.X^{i''}\rangle \approx \frac{2\sqrt{2}}{\pi} \frac{1-8\vba^6}{1+8\vba^6} \vba (1-\vba^2)\,.
\eea
In addition the constraint $\dot{\bf X}^2+{\bf X'}^2=1$ implies $\vba^2+\overline{t}^2=1$. The velocity-one scale model is used to evolve these quantities subject to the equations:
\bea
\frac{d\xi}{d\eta} = \vba^2 \mathcal{H} \xi+\frac{\tilde{c} }{2 }\vba\,,\qquad \frac{d\vba}{d\eta}=(1-\vba^2)\Big(\frac{\tilde{k}}{\xi}-2 \mathcal{H} \vba\Big)
\eea
where $\mathcal{H}=d\ln{a}/d\eta$, $\tilde{k}=({2\sqrt{2}}/{\pi}) ({1-8\vba^6})/({1+8\vba^6}) $, and $\tilde{c}$ denotes a constant energy loss rate (with value $0.23$ to agree with CMBACT).  

While the correlation functions described in equation~\eqref{eq:correls} are equal time correlators, we shall also require the unequal time correlator $\langle [\bXs(\sigma,\eta)- \bXs(\sigma',\eta')]^2 \rangle$. For small time differences and small spatial separations (such that $|\eta - \eta'|\equiv|\eta_-|<\xi$ and $|\sigma_-|<\xi$), we may approximate this quantity as \cite{Regan:2009hv}
\begin{align}
\langle [\bXs(\sigma,\eta)- \bXs(\sigma',\eta')]^2 \rangle&\approx \Gamma(\sigma_-,\eta_+)+V(0,\eta_+) \eta_-^2\,,
\end{align}
where $\eta_+=(\eta+\eta')/2$.

\subsection{Gott-Kaiser-Stebbins (ISW) effect}
As derived in \cite{Hindmarsh:1993pu}, the ISW contribution resulting from a photon moving with 4-momentum $p^\mu$ and intersecting the past-light cone of the string, is given by
\bea
\nabla_\perp^2 \frac{\delta T}{T}=-8\pi G\mu \int d\sigma {\bf u}.{\bf\nabla}_{\perp}\delta^{(2)}({\bf x} -{\bf X})\,,
\eea 
where ${\bf u}={\bf \dot{X}}-(X'.\hat{p}/\dot{X}.\hat{p}){\bf X}'$, and where the quantities expressed are evaluated at the light cone crossing time $\eta_r=\eta+z-\hat{\bf p}\cdot {\bf X}(\sigma,\eta_r)$, and $\delta^{(2)}$ represents the two dimensional Dirac delta function. Using the notation $\delta\equiv\delta T/T$ we write the expression in Fourier space in the form,
\bea
k^2 \delta_k(\eta_r) = - i 8\pi G\mu k^A \int d\sigma u^A(\sigma,\eta_r) e^{i {\bf k.X} (\sigma,\eta_r)}\,,
\eea
where $A=1,2$ runs over the transverse coordinates. To compute the total contribution to the ISW effect it is necessary to integrate over all contributions from the last-scattering surface to today.

The power spectrum, $dP$, and bispectrum, $dB$, contributions due to strings crossing the light cone at $\eta_r$ are given respectively by
\bea \label{eq:powerflat}
\langle \delta_{\bf k_1}(\eta_r)\delta_{\bf k_2}^*(\eta_r)\rangle &=& (2\pi)^2 \delta_D^{(2)}(\bk_1- \bk_2) dP(k_1,\eta_r)\,,\\
\langle \delta_{\bk_1}(\eta_r)\delta_{\bk_2}(\eta_r)\delta_{\bk_3}(\eta_r)\rangle&=&(2\pi)^2\delta_D^{(2)}(\bk_1+\bk_2+\bk_3) dB(k_1,k_2,k_3;\eta_r)\,,\label{eq:bispflat}
\eea
where $ \delta_D^{(2)}$ is the two dimensional Dirac delta function (which imposes the triangle condition in the flat sky limit in the case of the bispectrum).
Since the formulae expressed are already expressed in the flat sky one may relate the power spectrum and bispectrum to their CMB equivalents using the Limber approximation \cite{2000ApJ...530...36B}, such that we expect the ISW expression to be accurate for angular scales $\ell\gtrsim 100$. More explicitly, the angular power spectrum and angular bispectrum contributions due to the string with light-cone crossing time $\eta_r$ in the form 
\bea\label{eq:powISW}
C_\ell(\eta_r) =dP(k^*,\eta_r)/x_r^2\,,\qquad b_{\ell_1 \ell_2 \ell_3}(\eta_r)=dB(k^*_1,k^*_2,k^*_3)/x_{r}^4\,,
\eea
where $x_{r}=\eta_0-\eta_{r}$, and $k^*_i=\ell_i/x_r$.

\subsection{Contributions from Last Scattering}\label{sec:lss}
Decomposing the temperature fluctuation in terms of spherical harmonics, $\frac{\Delta T}{T}({\hat{\bf n}})=\sum_{lm} a_{lm}Y_{lm}(\hat{\bf n})$, we may express the coefficients $a_{lm}$ due to the contributions from last scattering, \eqref{eq:tempflucs}, as
\bea
a_{\ell m}\supseteq \int d\hat{\bn}Y_{\ell m}^*(\hat{\bn}) \int \frac{d^3 \bk}{(2\pi)^3} \Big[\frac{\delta_r(\bk)}{4}+ \hat{\bn}\cdot{\bf v}_r(\bk)\Big] e^{i \bk.\hat{\bn} x_{\rm dec}}
\eea
where $x_{\rm dec}=\eta_0-\eta_{\rm dec}$. The velocity term may be written in terms of its scalar and vector decomposition as
\bea
{\bf v}_r(\bk) = i\hat{\bk} v^{S}_r(\bk) + {\bf v}_r^V(\bk)\,,
\eea
where ${\bf v}_r^V.\bk =0$. As we shall note later, the vector part ${\bf v}^V_r$ may be neglected. Therefore, the spherical decomposition of the temperature fluctuation may be further 
expanded to give
\bea\label{eq:alm}
a_{\ell m}= 4\pi i^\ell\int \frac{d^3 \bk}{(2\pi)^3} \Big[\frac{\delta_r(\bk)}{4} j_\ell(k x_{\rm dec}) + v^S_r(\bk)j_\ell'(k x_{\rm dec})\Big] Y_{\ell m}^*(\hat{\bk})\,,
\eea
where $j_\ell'(x) = d j_{\ell}(x)/dx$, with $j_\ell$ denoting the spherical bessel function.

The governing equations we shall use assume a tightly-coupled photon-baryon fluid together with a cold dark matter component\footnote{We shall neglect the effect of neutrinos in our analysis} \cite{2000astro.ph.12205P}. The scale factor, $a(\eta)$, evolves according to the Friedmann equation,
\bea\label{eq:friedmann}
\frac{d a}{d\eta} = H_0\sqrt{ {\Omega_m}{a}  +\Omega_r +\frac{\Omega_\Lambda}{a^4}    }\,.
\eea
We normalise the scale factor to satisfy $a(\eta_0)=1$ today. As described in \cite{2000astro.ph.12205P} this results in the values at matter-radiation equality for the conformal time and scale factor $\eta_{\rm eq}=16.31/(\Omega_m h^2) {\rm Mpc}$, $a_{\rm eq}=1/(23219\Omega_m h^2)$,  with the conformal time today given by $\eta_0\approx13824 {\rm Mpc}$. In addition we assume that the conformal time at last scattering (as defined using the peak of the visibility function) is $\eta_{\rm dec}\approx 277 {\rm Mpc}$. We use the {\it Planck} \cite{2015arXiv150201589P} parameters, $\Omega_m=0.32$, $h=0.67$, $\Omega_\Lambda=0.68$, with $\Omega_r=\Omega_m a_{\rm eq}$. 

The linear perturbation equations governing the evolution of radiation ($r$) and cold dark matter ($c$) up until recombination are given by 
\bea
\ddot{\delta_c}+\frac{\dot{a}}{a}\dot{\delta}_c-\frac{3}{2}\left(\frac{\dot{a}}{a}\right)^2 \left(\frac{a\delta_c +(2+R)a_{\rm eq}\delta_r}{a+a_{\rm eq}+\frac{\Omega_\Lambda}{\Omega_c}a^4 }\right)&=&4\pi G\Theta_+\,,\nonumber\\
\dot{\delta_r}+\frac{4}{3}{\bf\nabla}\cdot{\bf v}_r -\frac{4}{3}\dot{\delta_c}&=&0\,,\\
\dot{\bf v}_r +\frac{\dot{a}}{a}\frac{R}{1+R}{\bf v}_r+\frac{1}{4+4R}{\bf\nabla}\delta_r&=&{\bf0}\,,
\eea
where we remind the reader that overdot represents differentiation with respect to conformal time, and where $\Theta_+=\Theta_{00}+\Theta_{i i}$ is the source term due to the string energy-momentum tensor. The variable, $R$, is defined as the ratio of the perturbation of the baryon ($b$) density to the photon density and is given by $R\equiv\delta \rho_B/\delta \rho_\gamma=3\rho_B/4\rho_\gamma\approx 0.6(\Omega_b h^2/0.02)(a/10^{-3})$. Since the vector component of ${\bf v}_r$, i.e. ${\bf v}^V_r$ is not sourced, we neglect this component and consider only the scalar part, $v^S_r$. These equations are solved using a Green's function technique, with
\bea
\ddot{\GG}_c+\frac{\dot{a}}{a}\dot{\GG_c}-\frac{3}{2}\left(\frac{\dot{a}}{a}\right)^2 \left(\frac{a\GG_c +(2+R)a_{\rm eq}\GG_r}{a+a_{\rm eq}+\frac{\Omega_\Lambda}{\Omega_c}a^4 }\right)&=&4\pi G\Theta_+\,,\nonumber\\
\dot{\GG_r}-\frac{k}{3}{\bar{\GG}}_{v_r} -\frac{4}{3}\dot{\GG_c}&=&0\,,\nonumber\\
\dot{\bar{\GG}}_{v_r}+\frac{\dot{a}}{a}\frac{R}{1+R}\bar{\GG}_{v_r} +\frac{k}{1+R}\GG_r&=&0\,,
\eea
subject to the initial conditions $\GG_c=0=\GG_r=\bar{\GG}_{v_r}, \dot{\GG}_c=1=3\dot{\GG}_r/4$ at $\eta=\eta'$ and $\GG_N=0$ for $\eta<\eta'$ (for $N\in\{r,s\}$). We have normalised the Green's functions such that the solutions for $\delta_r/4$ and $v_r^S$ are given, respectively, by
\bea
\frac{\delta_r}{4}(\bk,\eta,\eta_i)=\pi G\int_{\eta_i}^{\eta} \GG_r(k;\eta,\eta') \Theta_+(\bk,\eta') d\eta'\,, \quad v_r^S(\bk,\eta,\eta_i)=\pi G\int_{\eta_i}^{\eta} \bar{\GG}_{v_r}(k;\eta,\eta') \Theta_+(\bk,\eta') d\eta'\,.\nonumber
\eea
\para{Compensation factor} One particular issue with this treatment is that the compensating under-density created due to the formation of the topological defects is not included. While a more involved treatment may account for such effects exactly, a simple prescription was described in \cite{Wu:1998mr} is the use of a compensating factor, $\gamma_c$ given by the kernel
\bea
\gamma_c(k,\eta)=\frac{1}{1+k_c^2(\eta)/k^2}\,,
\eea
where $k_c(\eta)\eta$ interpolates between $\sqrt{6}$ in the radiation era and $\sqrt{18}$ in the matter era. 
\para{Damping factors} Our treatment of recombination has thus far assumed an instantaneous decoupling. In reality the last scattering surface has a finite thickness resulting in a diffusion damping
of photons. One may account for these using a damping envelope \cite{1996ApJ...471..542H}, $\mathcal{D}_\gamma(k)\approx e^{-(k/k_\gamma)^{m_{\gamma}}}$, where for the parameters used in this paper we may set $k_\gamma=0.109 \,{\rm Mpc}^{-1}$ and $m_\gamma=1.45$.

We may summarise these effects in the intrinsic and Doppler terms by writing,
\bea\label{eq:intrinsicdoppler}
\frac{\delta_r}{4}(\bk,\eta,\eta_i)&=&\pi G\int_{\eta_i}^{\eta} \mathcal{D}_\gamma(k)\GG_r(k;\eta,\eta') \Theta_+(\bk,\eta')\gamma_c(k,\eta')d\eta'   \,, \nonumber\\
 v_r^S(\bk,\eta,\eta_i)&=&\pi G\int_{\eta_i}^{\eta} \mathcal{D}_\gamma(k)\bar{\GG}_{v_r}(k;\eta,\eta') \Theta_+(\bk,\eta') \gamma_c(k,\eta')d\eta'  \,.
\eea
We simplify notation and set 
\bea\label{eq:intrinsicdopperGreens}
\GG^{i}(k;\eta,\eta') &\equiv& \GG_r (k;\eta,\eta') \gamma_c(k,\eta') \mathcal{D}_\gamma(k)\,, \nonumber\\
\GG^{d}(k;\eta,\eta') &\equiv& \bar{\GG}_r (k;\eta,\eta') \gamma_c(k,\eta') \mathcal{D}_\gamma(k)\,,
\eea
with the superscripts $i$ and $d$ used to denote intrinsic and Doppler, respectively.
The advantage of the approach used, as described earlier, is that the calculation of the bispectrum becomes tractable. At the level of accuracy required for this paper, the approximations made are deemed to be adequate to assess the relative importance of the terms at last scattering to the total CMB bispectrum. In §\ref{sec:power} we shall provide a check of the qualitative and quantitative fit to the power spectrum using these approximations by making a comparison to that obtained using a full Boltzmann code treatment.

\section{Power Spectrum }\label{sec:power}
In this section we compute the CMB power spectrum due to a network of cosmic strings. We shall first present our computation of the ISW power spectrum, and, subsequently, our calculation of the power spectra due to the intrinsic and Doppler contributions from the last scattering surface. Due to the approximations made throughout this work, we expect that the power spectrum deviates on superhorizon scales on the last scattering surface. Nevertheless, our aim is to accurately model effects on scales of the correlation length of the string network (at last scattering) and below, since the signal to noise is dominated on those scales. We shall test the accuracy of our approximations by comparison to the power spectrum computed using CMBACT.

\subsection{ISW Power Spectrum}
In order to compute the total ISW power spectrum we integrate up all contributions from recombination to today. We shall assume that the contributions from strings at unequal times are uncorrelated. In the flat-sky approximation used to compute the ISW temperature fluctuations, the power spectrum contribution, $dP$,  due to strings crossing the light cone at $\eta$ is given by equation \eqref{eq:powerflat}.
The total angular power spectrum, $C_\ell$ is then given by (c.f. equation~\eqref{eq:powISW})
\bea
\ell^2 C_\ell= \int_{\eta_{\rm start}}^{\eta_0} d\eta k^{*2} \frac{dP(k^*,\eta)}{d\eta}\,, 
\eea
where $k^*=\ell/(\eta_0-\eta)$. On sufficiently small angular scales $\eta_{\rm start}$ is given by $\eta_{\rm dec}$, while more generally we shall define $\eta_{\rm start}$ to ensure validity of the Gaussian model. More precisely the Gaussian model is assumed to be valid for length scales, $k^{-1}$, satisfying $k^{-1}\lesssim \xi\equiv \alpha \eta$. Therefore we define, $\eta_{\rm start}$, such that
\bea
\eta_{\rm start}={\rm max}(\eta_{\rm dec},(k\alpha)^{-1})\,.
\eea 
Our calculation of the power spectrum requires the calculation of the two point functions of the string network functions, as described in \cite{Hindmarsh:1993pu,Hindmarsh:2009qk,Regan:2009hv} resulting in the power spectrum contribution due to strings crossing at $\eta$  given by
\bea
dP(\eta)&=&\frac{(8\pi G\mu)^2}{\mathcal{A} k^4}\int d\sigma d\sigma' k^A k^B\langle u^A(\sigma,\eta) u^B(\sigma',\eta) e^{i \bk.({\bf X}(\sigma)-{\bf X}(\sigma'))}\rangle \nonumber\\
&=&(8\pi G\mu)^2 \frac{d\mathcal{L}}{\mathcal{A}}\frac{1}{2 k^2}\int_{-\xi}^{\xi} d\sigma_- U(\sigma_-,\eta)e^{-{\Gamma}(\sigma_-,\eta) k^2/6}\,,
\eea
where  $\langle u^A(\sigma) u^B(\sigma')\rangle =(\delta_{AB}/2)U(\sigma-\sigma')$, $k^2=k^A k^A$, $\mathcal{A}=(2\pi)^2\delta_D^{(2)}(0)$ is a formal area factor, and $\mathcal{L}$ represents the total string length of the network at $\eta$. Since the total length per unit volume of the string network is given by $1/\xi^2$, we infer that 
\bea
\frac{d\mathcal{L}}{\mathcal{A}d\eta}\approx\frac{c}{\alpha\eta^2}\,,
\eea
where $c$ is a constant which was deduced to be approximately unity \cite{Regan:2009hv}. We use the asymptotic small scale limit for the two point function $U(\sigma)\approx \bar{u}^2$, which is related to $\bar{v}$ (given by equation \eqref{eq:small_scale}) via
\bea
U(\sigma)\approx \bar{u}^2 \approx \frac{2}{3}\vba^2 +\frac{2}{9}(1-\vba^2)^2(1+\vba^2)\,.
\eea
The total angular power spectrum is therefore given by
\bea \label{eq:ISWpower}
\ell^2 C_{\ell} = (8\pi G\mu)^2 \int_{\eta_{\rm start}}^{\eta_0} d\eta \frac{1}{\eta^2}\frac{\sqrt{6\pi}\bar{u}^2}{2\alpha \bar{t}}\frac{1}{k^*} {\rm erf}\left(\frac{k^*\bar{t} \alpha \eta}{\sqrt{6}}\right)\,, 
\eea
where we simplify notation by suppressing the implicit time dependence of the quantities $k^*\equiv\ell/(\eta_0-\eta)$ and of the string network parameters $\bar{u}, \bar{t}, \bar{v},{\alpha}$. 

\subsection{Intrinsic and Doppler Power Spectra}\label{sec:powerlss}
The Fourier space power spectrum of a three dimensional quantity, $\delta(\bk)$, is given by
\bea
\langle \delta(\bk_1) \delta(\bk_2)\rangle = (2\pi)^3 \delta_D^{(3)}(\bk_1+\bk_2+\bk_3) P(k)\,,
\eea
where $\delta_D^{(3)}$ represents the three dimensional Dirac delta function. We shall denote by $P^{i}$ the power spectrum of the intrinsic perturbation $\delta_r/4$, and by $P^{d}$ the power spectrum of the Doppler perturbation $v_r^S$. In addition we shall denote the cross spectrum by $P^{i \times d}$ (with $P^{i\times i}\equiv P^i$ and $P^{d\times d}\equiv P^d$). Following the Green's function technique described in §\ref{sec:lss}, the respective power spectra are given by
\bea
P^{a\times b}(k) =(\pi G)^2 \iint_{\eta_i}^{\eta_{\rm dec}}d\eta_1 d\eta_2 \GG^{a}(k;\eta_{\rm dec},\eta_1)\GG^{b}(k;\eta_{\rm dec},\eta_2)\frac{\langle \Theta_+(\bk,\eta_1) \Theta_+^*(\bk,\eta_2)\rangle}{\mathcal{V}}\,,
\eea
where $a,b\in\{i,d\}$, and $\mathcal{V}\equiv (2\pi)^3 \delta_D^{(3)}(0)$ represents a formal volume factor, and $\eta_i$ denotes the initial time (taken in this paper to be $0.01\, {\rm Mpc}$). The unequal time correlator (UETC), $\langle \Theta_+\Theta_+^*\rangle/\mathcal{V}$, is given in the Gaussian model by \cite{2015JCAP...03..008R}
\bea
\frac{\langle \Theta_+(\bk,\eta_1) \Theta_+^*(\bk,\eta_2)\rangle}{\mathcal{V}}=4\mu^2\frac{7}{6}\frac{\sqrt{6\pi}\vba^4}{\alpha^2 \bar{t}} \frac{1}{k \eta_1 \eta_2}{\rm erf}\left(k\bar{t}\alpha\sqrt{\frac{\eta_1\eta_2}{6}}\right){\rm exp}\left(-\frac{k^2\vba^2\eta_{12}^2}{6}\right)\frac{2(\eta_1 \eta_2)^{3/2}}{\eta_1^3+\eta_2^3}\,,\nonumber
\eea
where $\eta_{12}\equiv\eta_1-\eta_2$. The final factor was advocated in \cite{2015JCAP...03..008R} to account for the super-horizon behaviour of the UETC.

The angular power spectrum, $C_\ell$, is defined by
\bea\label{eq:cell}
\langle a_{\ell_1 m_1} a_{\ell_2 m_2}^*\rangle = C_{\ell_1}\delta_{\ell_1 \ell_2}\delta_{m_1 m_2}\,,
\eea
and may be inferred from equation~\eqref{eq:alm} to be of the form
\bea
C_\ell^i &=& \frac{2}{\pi} \int P^i(k) (j_\ell(k x_{\rm dec}))^2 k^2 dk\,, \\
C_\ell^d &=& \frac{2}{\pi} \int P^d(k) (j'_\ell(k x_{\rm dec}))^2 k^2 dk\,,\\
C_\ell^{i\times d} &=& \frac{2}{\pi} \int P^{i \times d}(k) j_\ell(k x_{\rm dec}) j'_\ell(k x_{\rm dec}) k^2 dk\,.
\eea

\subsection{Comparison to CMBACT}
\begin{figure}
\centering
\vspace{0.25cm}
\hspace{0.1cm}
{\includegraphics[width=0.8\linewidth]{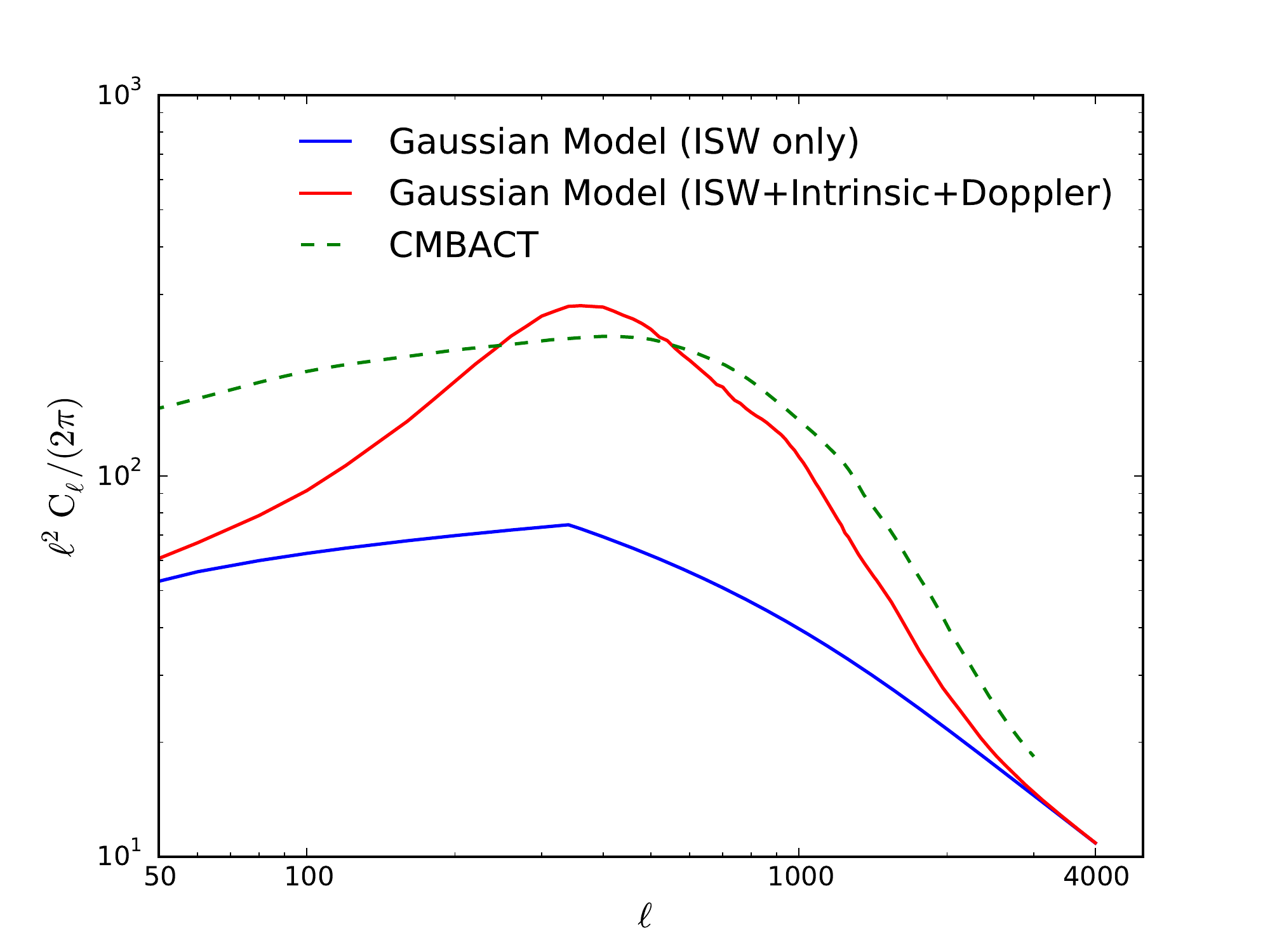}}
\caption{Comparison between the angular power spectrum obtained using the full Boltzmann code (using CMBACT) and that obtained using the Gaussian model approach used in this paper. The quantity $\ell^2 C_\ell$ is plotted in units of the string tension, $G\mu$. We plot the total contribution including the integrated Sachs Wolfe, intrinsic and Doppler effects, as well as the contribution due to the ISW term alone. The Gaussian model is seen to provide a reasonably accurate fit on angular scales $\ell\gtrsim 200$.}
\label{fig:PowerSpec}
\end{figure}

The various approximations made have been done so since the scope of this paper is to provide an estimate for the relative magnitude of the bispectrum due to last scattering surface effects compared to that due to the integrated Sachs Wolfe effect. We compute the ISW, intrinsic and Doppler terms using the {\it{Planck}} parameters as given below equation~\eqref{eq:friedmann} and using the initial conditions for the velocity-one scale model as used by CMBACT, i.e. $\vba(\eta_i)=0.65$, $\alpha(\eta_i)=0.15$, with $\eta_i=0.01\,{\rm Mpc}$. We have set the string tension to $G\mu=10^{-6}$ for reference. In order to assess the impact of the various approximations we compare our computation to that obtained using a full Boltzmann treatment, including the effects of baryons and photon diffusion exactly. CMBACT uses such a treatment, with the UETCs of the string network obtained using simulations of the unconnected segment model. While the cross term between the intrinsic and Doppler terms is found to be negligible, it is found that on length scales around the correlation length at last scattering (corresponding to angular scales $\ell\sim 400$) the contributions due to last scattering surface effects  exceed those due to the total ISW effect. In Figure~\ref{fig:PowerSpec} we plot the comparison between CMBACT, the total obtained using the Gaussian model approach used in this paper (including ISW, intrinsic and Doppler effects), and also, for reference, plot the ISW only contribution. We observe that on scales for angular scales $\ell\gtrsim 200$ the fit to the power spectrum obtained using the approach adopted in this paper replicates both qualitatively and quantitatively the features of the power spectrum obtained using the full Boltzmann code. The discrepancy for scales $\ell \lesssim 200$ is possibly due to the neglect of super-horizon physics and compensation effects in the integrated Sachs Wolfe term, which - calculated using a flat-sky formalism - could only be expected to be accurate for angular scales well above $\ell\sim100$. We conclude that our approach is precise to a sufficient level of accuracy to faithfully estimate the signal to noise improvement to the bispectrum that might be obtained via the full inclusion of terms from the last scattering surface.

\section{Bispectrum }\label{sec:bisp}
The CMB bispectrum is given by the three point function of the the spherical harmonic coefficients of the temperature fluctuations,
\bea\label{eq:bispell}
\langle a_{\ell_1 m_1}a_{\ell_2 m_2}a_{\ell_3 m_3}\rangle = G^{\ell_1 \ell_2 \ell_3}_{m_1 m_2 m_3} b_{\ell_1 \ell_2 \ell_3}\,,
\eea
where the $G^{\dots}_{\dots}$, representing the Gaunt integral, imposes the triangle condition in multipole space and is given by
\bea
G^{\ell_1 \ell_2 \ell_3}_{m_1 m_2 m_3} =\int d\hat{\bn} Y_{\ell_1 m_1}(\hat{\bn})Y_{\ell_2 m_2}(\hat{\bn})Y_{\ell_3 m_3}(\hat{\bn})\,.
\eea
One is primarily interested in the signal to noise measure in order to assess the detectability or otherwise of this signal, given by
\bea\label{eq:signoise}
\left(\frac{S}{N}\right)^2 = \frac{f_{\rm sky}}{6}\sum_{\ell_i} \frac{b_{\ell_1 \ell_2 \ell_3}^2 h_{\ell_1 \ell_2 \ell_3}^2}{C^{\rm tot}_{\ell_1}C^{\rm tot}_{\ell_2}C^{\rm tot}_{\ell_3}}\,,
\eea
where $h_{\ell_1 \ell_2 \ell_3} \equiv \sqrt{\dfrac{(2\ell_1+1)(2\ell_2+1)(2\ell_3+1)}{4\pi}}\bigg( \begin{array}{ccc}
\ell_1 & \ell_2 & \ell_3 \\
0 & 0 & 0 \end{array} \bigg)$, with the second factor denoting the Wigner $3$-j symbol (which, inherited from the Gaunt integral, imposes the triangle condition), and $f_{\rm sky}$ denotes the fractional sky coverage of the survey considered. The power spectrum $C^{\rm tot}_\ell$ is the CMB power spectrum, observed to show high consistency with the inflationary paradigm. In order to compute the signal to noise accurately we include the effect of beam and noise effects in our estimates, as described in §\ref{sec:results}. Computing the signal to noise for a reference value of $G\mu$, we note that $1\sigma$ error bars on the quantity $(G\mu/10^{-6})^3$ is given by
\bea
\Delta (G\mu/10^{-6})^3 = \frac{1}{\sqrt{\left({S}/{N}\right)^2_{G\mu=10^{-6}}}}\,.
\eea
In this section we shall outline how the CMB bispectrum of the ISW, intrinsic and Doppler terms are computed in the Gaussian model framework described in §\ref{sec:method}.

\subsection{ISW Bispectrum}
As in the case of the power spectrum, we compute the ISW bispectrum by integrating up all contributions between last scattering and today, assuming that contributions from unequal times are uncorrelated. The computation is carried out using the flat-sky approximation, with the bispectrum contribution, $dB$ due to strings crossing the light cone at $\eta$ given by equation~\eqref{eq:bispflat}. The Limber approximation, \eqref{eq:powISW}, may then used to give the CMB bispectrum result. Explicitly we compute,
\bea
b_{\ell_1 \ell_2 \ell_3}=\int_{\eta_{\rm start}}^{\eta_0} d\eta \frac{1}{(\eta_0-\eta)^4} \frac{dB(k_1^*,k_2^*,k_3^*;\eta)}{d\eta} \,,
\eea
where we shall define $\eta_{\rm start}$ later in the section to ensure validity of the Gaussian model for the length scales integrated over, $k_i^*=\ell_i/(\eta_0-\eta)$. In the Gaussian model we may readily compute the contribution, $dB$, in a similar fashion to the power spectrum as 
\bea
dB(k_1,k_2,k_3;\eta)=i\frac{(8\pi G\mu)^3}{\mathcal{A} k_1^2 k_2^2 k_3^2}\int d\sigma_1d\sigma_2 d\sigma_3 \langle k_1^A k_2^B k_3^C u^A (\sigma_1)u^B (\sigma_2) u^C  (\sigma_3)e^{i D}\rangle\,,
\eea
where we set $k_3=|\bk_1+\bk_2|$, $\mathcal{A}=(2\pi)^2\delta_D^{(2)}({\bf 0})$, $A,B,C\in \{1,2\}$ and $D=\sum_{i=1}^3\bk_i.{\bf X}_i$. Utilising the Gaussian model functions described in §\ref{eq:gaussmodel} this equation results in the expression \cite{Hindmarsh:2009qk,Regan:2009hv}
\bea \label{eq:ISWbisp}
\frac{dB(k_1,k_2,k_3;\eta)}{d\eta}=-\frac{(8\pi G\mu)^3 }{ \eta^3 (k_1 k_2 k_3)^2}\frac{c_0 \bar{u}^2}{8\alpha^2}\int && d\sigma_{12}d\sigma_{13} \left[k_1^2 \kappa_{23}\sigma_{23}^2 +k_2^2 \kappa_{13}\sigma_{13}^2 +k_3^2 \kappa_{12}\sigma_{12}^2 \right]\nonumber\\
&&\times {\rm exp}\left(-\frac{\sum_{i<j}\bar{t}^2 \kappa_{ij} \sigma_{ij}^2}{6}\right)\,,
\eea
where $\sigma_{i j}\equiv \sigma_i-\sigma_j$, and $\kappa_{ij}=-\bk_i.\bk_j$. In \cite{Hindmarsh:2009qk} it was shown that this integral can be computed analytically in a sufficiently small angle limit, which may be deduced from \cite{Regan:2009hv} to be given by the condition $(\sqrt{k_1^2 k_2^2 -\kappa_{12}^2}/k_i) \xi \gtrsim 4/\bar{t} $, noting that the area of the triangle defined by wavenumbers $k_1,k_2,k_3$ is given by $\sqrt{k_1^2 k_2^2 -\kappa_{12}^2}/2$. While one may produce analytic approximations to the bispectrum on smaller scales, the signal is suppressed, so we instead consider only time scales satisfying this condition, i.e. we impose the start time of our integration to be given by
\bea\label{eq:etastart}
\eta_{\rm start}={\rm max}\left(\eta_{\rm dec}, \frac{4}{\alpha \bar{t}}\frac{k_i}{\sqrt{k_1^2 k_2^2 -\kappa_{12}^2}} \right)\,.
\eea
In $\ell$ space this translates to the condition $\eta_{\rm start}\approx {\rm max}(\eta_{\rm dec},\eta_0 (30/\hat{\ell}_i) )$, where $\hat{\ell}_i=\ell_1 \ell_2 |\sin \theta_{12}|/\ell_i$, where $\theta_{12}$ is the angle between the triangle sides of length $k_1$ and $k_2$. Thus for triangles that are too small or angular scales which are too small such that $\hat{\ell}_i$ there is no contribution to the bispectrum integral. This is to be expected, since our use of the flat-sky approximation is not anticipated to be valid for angular scales $\ell \lesssim 100$.

The bispectrum contribution is then given by
\bea\label{eq:contrib}
\frac{dB(k_1,k_2,k_3;\eta)}{d\eta}=-\frac{(8\pi G\mu)^3 }{ \eta^3 (k_1 k_2 k_3)^2}\frac{9\pi c_0 \bar{u}^2}{4\bar{t}^4\alpha^2} \frac{k_1^4 \kappa_{23}+k_2^4 \kappa_{13}+k_3^4 \kappa_{12}}{(\kappa_{12}\kappa_{13}+\kappa_{12}\kappa_{23}+\kappa_{13}\kappa_{23})^{3/2}} \,.
\eea
One may readily compute the CMB ISW bispectrum from cosmic strings for particular configurations using the above expressions. From equations~\eqref{eq:etastart} and \eqref{eq:contrib} one may deduce that in the equilateral configuration for which $\eta_{\rm start}\approx 30 (14000/\ell)$, the ISW bispectrum will scale as $\ell^{-4}$ for $\ell\lesssim 1400$, and as $\ell^{-6}$ for larger $\ell$ values. In order to compute the signal to noise, \eqref{eq:signoise}, one requires the bispectrum for all $\ell_i\in [2,2000]$. We achieve this by computing a grid of values and fitting to a basis in order to smoothly interpolate to arbitrary configurations. The method used to achieve this fitting is described in Appendix~\ref{sec:AppA}.

\subsection{Intrinsic and Doppler Bispectra}
The Fourier space bispectrum is defined as
\bea
\langle \delta(\bk_1)\delta(\bk_2)\delta(\bk_3)\rangle = (2\pi)^3 \delta_D^{(3)}(\bk_1+\bk_2+\bk_3) B(k_1,k_2,k_3)\,,
\eea
where the Dirac delta condition (due to statistical homogeneity) imposes the triangle condition, such that the bispectrum may be expressed using only the wavenumbers, $k_i$.
The bispectrum for the intrinsic ($i$) and Doppler ($d$) terms are denoted $B^{(a)}$ for $a\in \{ i,d\}$. 
Using equation~\eqref{eq:alm} and re-writing the Dirac delta function as $(2\pi)^3 \delta_D^{(3)}(\bk)=\int d^3\bx {\rm exp}(i \bk\cdot \bx)$, the CMB bispectrum for the intrinsic and Doppler terms is inferred to be given by 
\bea\label{eq:CMBbisp}
b_{\ell_1 \ell_2 \ell_3}^{(a)} = \left(\frac{2}{\pi}\right)^3 \int dx x^2 \int dk_1 dk_2 dk_3 (k_1 k_2 k_3)^2 B^{(a)}(k_1,k_2,k_3) \Pi_{i=1}^3 \left[ j^{(a)}_{\ell_i}(k_i x_{\rm dec}) j_{\ell_i}(k_i x)\right]\,,\nonumber\\
\eea
where $a\in \{ i,d\}$ labels the intrinsic or Doppler terms, and we introduce the symbols $j_\ell^{(i)}\equiv j_{\ell}$ and $j_\ell^{(d)}\equiv j'_{\ell}$ to simplify notation. The line of sight integral, $\int dx$, is carried out between last scattering and today, while we choose the upper limit of the integration with respect to the wavenumbers $k_i$ to ensure convergence (more explicitly we choose $k_{\rm max}=0.05\,{\rm Mpc}^{-1}$). Computation of the CMB bispectrum is, in general, a numerically prohibitive task taking $\mathcal{O}(10^{15})$ computations for $Planck$ resolution of $\ell_{\rm max}=2000$, unless the Fourier space bispectrum $B^{(a)}$ can be expressed in a separable form. In Appendix~\ref{sec:AppA}, we describe a mechanism for performing a separable decomposition, given a grid of values. The mechanism may also be used to decompose the ISW bispectrum. This allows for an efficient calculation of the CMB bispectrum in terms of a basis $Q_n$ of separable terms with coefficients $\bar{\alpha}_n^{(a)}$ for $a\in\{ i,d,\rm ISW\}$. The signal to noise may then be computed readily using equation~\eqref{eq:signoiseexpan} as with the correlations between the different terms.

In this section we outline the calculation of the intrinsic and Doppler Fourier bispectra. We neglect the mixed terms which are expected to be negligible following our considerations of the power spectrum. Following the Green's function technique in §\ref{sec:lss}, and mirroring the steps used in §\ref{sec:powerlss} for the power spectrum, we calculate the bispectra according to
\bea
B^{(a)}(k_1,k_2,k_3)=(\pi G)^3 \iiint_{\eta_i}^{\eta_{\rm dec}} \Pi_{i=1}^3\left[d\eta_i \GG^a(k_i;\eta_{\rm dec},\eta_i)\right]\frac{\langle \Theta_+(\bk_1,\eta_1)\Theta_+(\bk_2,\eta_2)\Theta_+(\bk_3,\eta_3)\rangle}{\mathcal{V}}\,.\nonumber\\
\eea
The three point unequal time correlator is given in the Gaussian model by \cite{2015JCAP...03..008R}
\bea
\frac{\langle \Theta_+(\bk_1,\eta_1)\Theta_+(\bk_2,\eta_2)\Theta_+(\bk_3,\eta_3)\rangle}{\mathcal{V}}&&\approx \frac{\beta_0}{6\xi^2}\frac{1}{\sqrt{k_1^2 k_2^2-\kappa_{12}^2}}{\rm erf}\left(\frac{\bar{t} k_2 \xi}{\sqrt{6}}\right){\rm erf}\left(\frac{\bar{t} \sqrt{k_1^2-\kappa_{12}^2/k_2^2} \xi}{\sqrt{6}}\right)\nonumber
\\ &&\times {\rm exp}\left(\frac{\vba^2}{6}(k_2^2 \eta_{12}^2+(k_1^2-\kappa_{12}^2/k_2^2)\eta_{13}^2)\right)\frac{\eta_1^4}{\eta_2^2\eta_3^2} + 5\,{\rm perms\, of\,} k_i\nonumber \,,\\
\eea
where $\beta_0=8\mu^3 (29\pi/3)(\vba^6/\bar{t}^2)$ and $\eta_{ij}=\eta_i-\eta_j$. In the above formula we set $\xi=\xi(\eta_1)$ - assuming, without loss of generality, that $\eta_1<\eta_{2,3}$ -, with the  factors on the second line accounting for unequal times, $\eta_{ij}\neq 0$.

\section{Results and Discussion }\label{sec:results}
We evaluate the CMB bispectrum due to cosmic strings induced by the intrinsic, Doppler and ISW terms separately. We neglect cross-terms in our analysis. Our goal is to assess the comparative improvement achievable in signal to noise from an ISW only bispectrum - as considered by {\it Planck} - once last scattering effects are accounted for.

The computation is carried out using the formulae given in §\ref{sec:bisp}. We assume the cosmological parameters listed under equation~\eqref{eq:friedmann}. Our estimates for the signal to noise, \eqref{eq:signoise} require the calculation of the angular power spectrum $C_\ell^{\rm tot}$, including the effects of noise and beam effects. We firstly generate the inflationary CMB power spectrum\footnote{One should note that our calculation of the cosmic variance represented in the denominator of \eqref{eq:signoise} neglects the cosmic string contribution, which is known to contribute at a level of less than $3\%$ \cite{Ade:2013xla}.}, $C_\ell^{\rm infl}$, using CAMB \cite{Lewis:1999bs} and include {\it Planck}-realistic beam, $b_\ell$, and noise, $N_\ell$ by setting $C_\ell^{\rm tot}= C_\ell^{\rm infl} + {N_\ell}/{b_\ell^2}$ - using the prescription outlined in \cite{Baumann:2008aq} - where
\bea
b_\ell&=&{\rm exp}(-\ell (\ell+1) \sigma_{\rm beam}^2);\quad \sigma_{\rm beam}=\frac{\theta_b}{\sqrt{8\ln 2}};\quad N_\ell= \sigma_{\rm pix}^2 \Omega_{\rm pix};\quad \Omega_{\rm pix}=\frac{4\pi}{N_{\rm pix}}\,.
\eea
$\theta_b$ denotes the resolution of the beam, $N_{\rm pix}$ labels the number of pixels of area $\Omega_{\rm pix}$, required to cover the sky, with $\sigma_{\rm pix}^2$ giving the variance per pixel. In the case of a {\it Planck}-realistic experiment we use the parameters $\theta_b=7.1'$, $\sigma_{\rm pix}=2.2\times 10^{-6}$ and $\Omega_{\rm pix}=0.0349$. 

The bispectra $b_{\ell_1 \ell_2 \ell_3}^{\rm ISW}, B^{i}(k_1,k_2,k_3), B^{d}(k_1,k_2,k_3)$ are evaluated on a grid of $200^3$ values for $\ell_i\in [2,2000]$ and $k_i\in [10^{-4},0.5] {\rm Mpc}^{-1}$ (with the latter logarithmically spaced). We perform the separable decomposition of the shapes, as described in Appendix~\ref{sec:AppA}, and,  - in the case of the intrinsic and Doppler terms, - perform the integrations necessary to obtain the corresponding CMB shape. We finally obtain the coefficients $\bar{\alpha}_n^{(a)}$ of the basis expansion of the CMB shapes 
\bea
s^{(a)}_{\ell_1 \ell_2 \ell_3}\equiv b^{(a)}_{\ell_1 \ell_2 \ell_3}\frac{(2\ell_1+1)^{1/6}(2\ell_2+1)^{1/6}(2\ell_3+1)^{1/6}}{\sqrt{C_{\ell_1}^{\rm tot}C_{\ell_2}^{\rm tot}C_{\ell_3}^{\rm tot} }} =\sum_n \bar{\alpha}_n^{(a)} Q_n(\ell_1,\ell_2,\ell_3)\,,
\eea
where $a\in \{i,d,{\rm ISW} \}$, and $Q_n$ represents the 3D basis of separable shapes (see Appendix~\ref{sec:AppA}). In this paper we use $160$ partial waves, such that we achieve an accuracy of greater than $99\%$ accuracy in our decomposition. 

\begin{figure}
\centering
\vspace{0.25cm}
\hspace{0.1cm}
{\includegraphics[trim = 35mm  95mm 40mm 95mm, clip,width=0.48\linewidth]{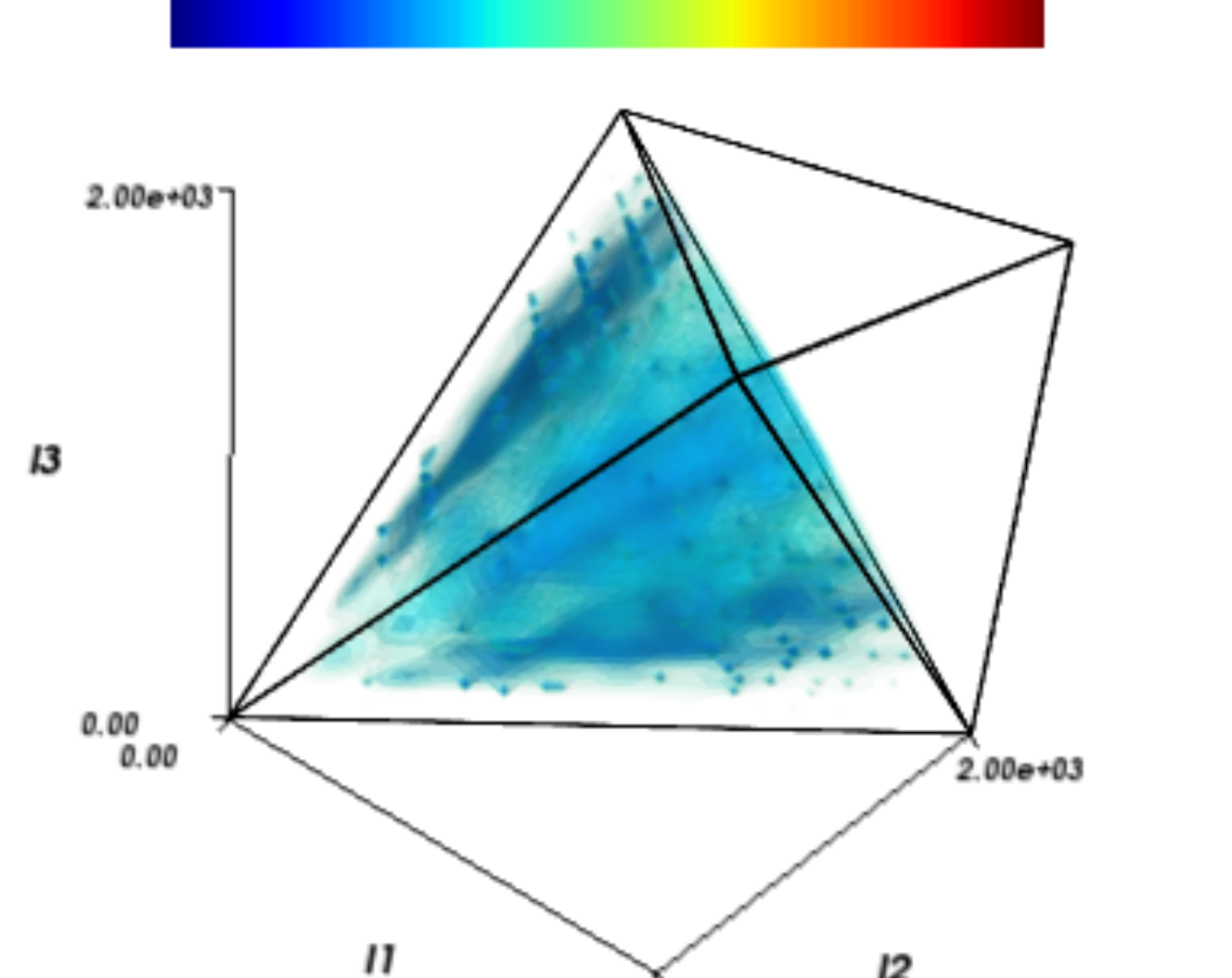}}
{\includegraphics[trim = 35mm  95mm 40mm 95mm, clip,width=0.48\linewidth]{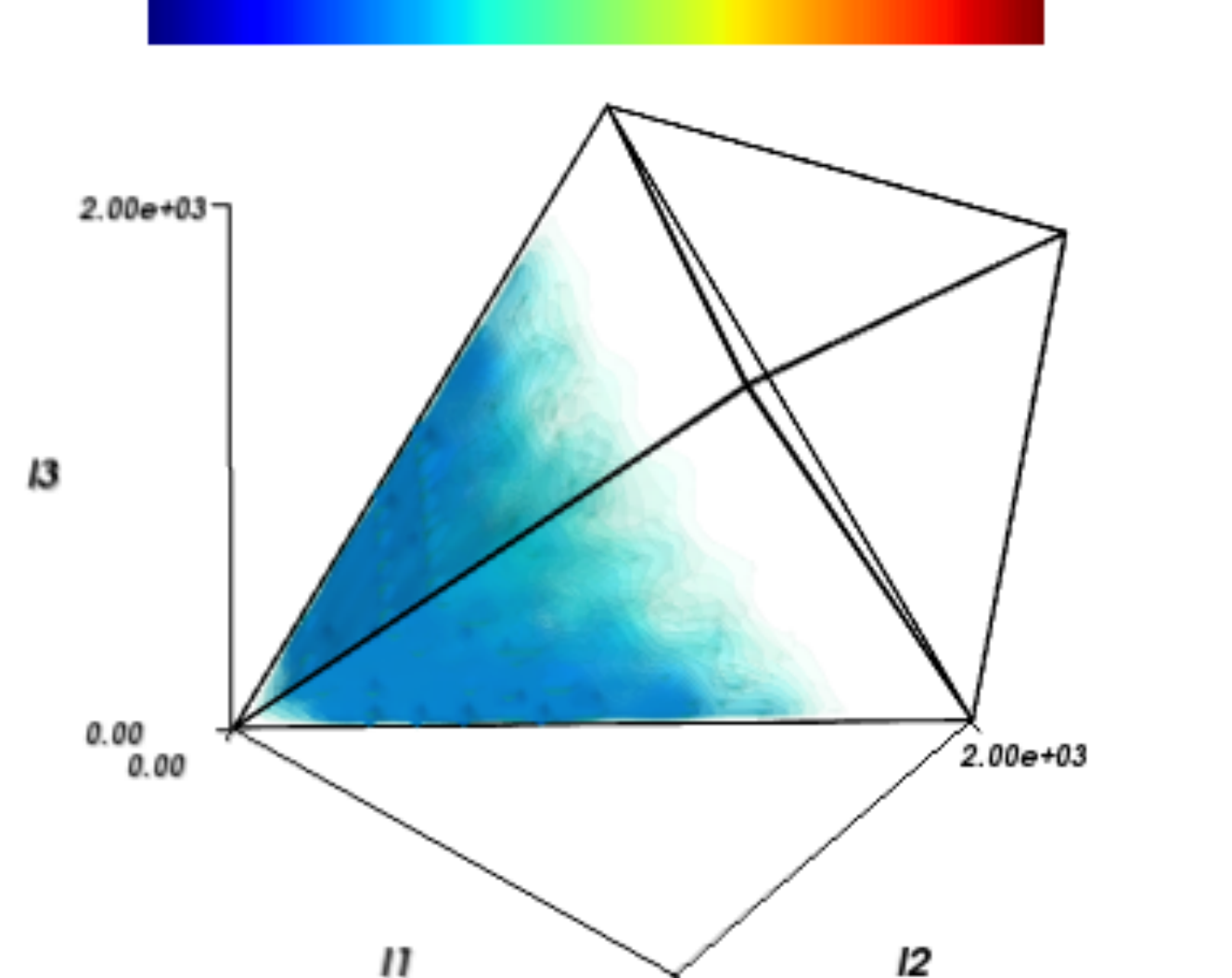}}
{\includegraphics[trim = 35mm  94mm 40mm 80mm, clip,width=0.48\linewidth]{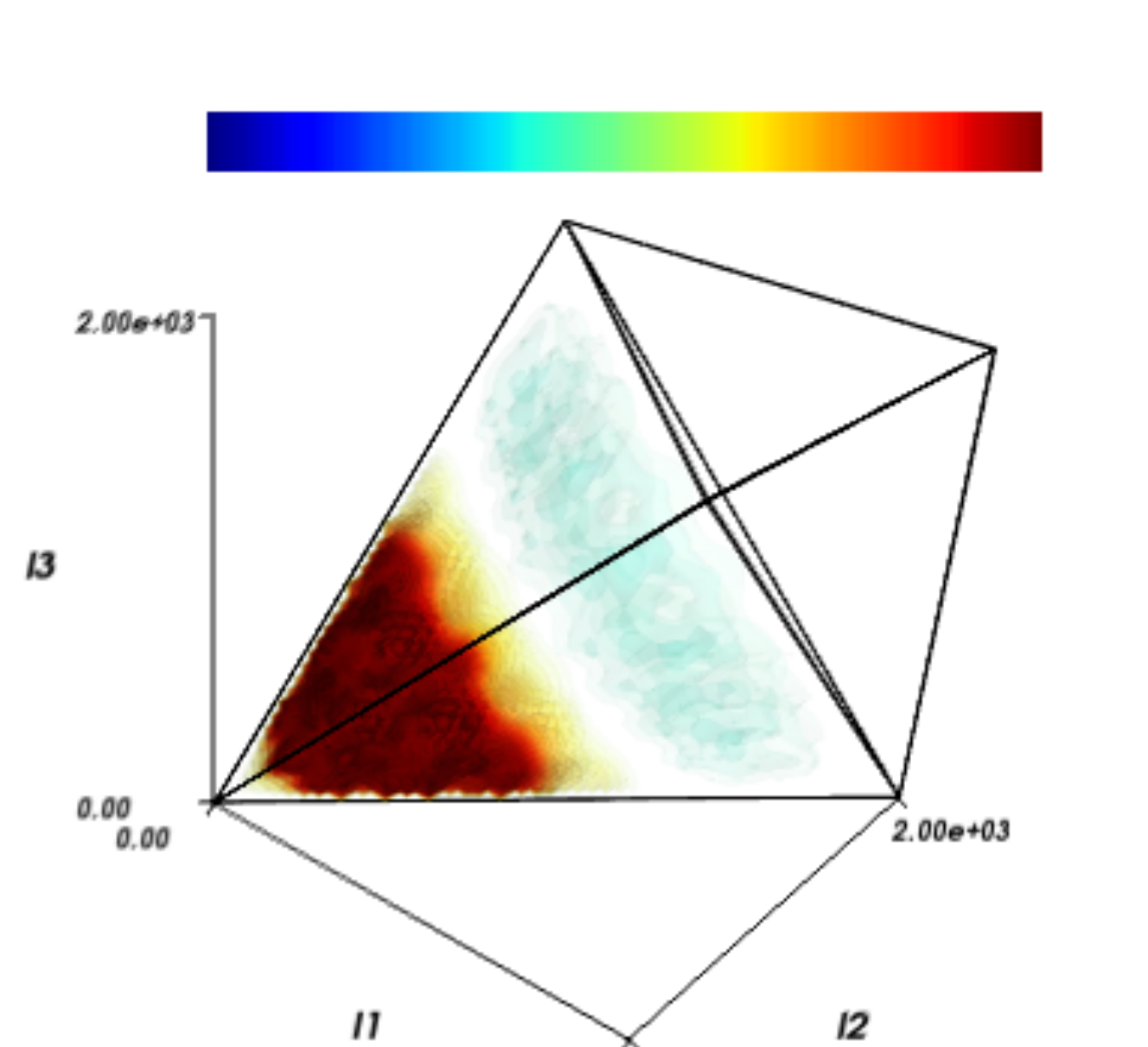}}
\caption{Clockwise from top left we plot the ISW, intrinsic and Doppler bispectra (normalised by the constant Sachs Wolfe bispectrum). The three shapes are clearly distinguishable with the Doppler term of opposite sign to the ISW and intrinsic bispectra. Note that the colour bars are not normalised equally.}
\label{fig:Bispec_3types}
\end{figure}

\begin{figure}
\centering
\vspace{0.25cm}
\hspace{0.1cm}
{\includegraphics[trim = 35mm  80mm 40mm 90mm, clip,width=0.48\linewidth]{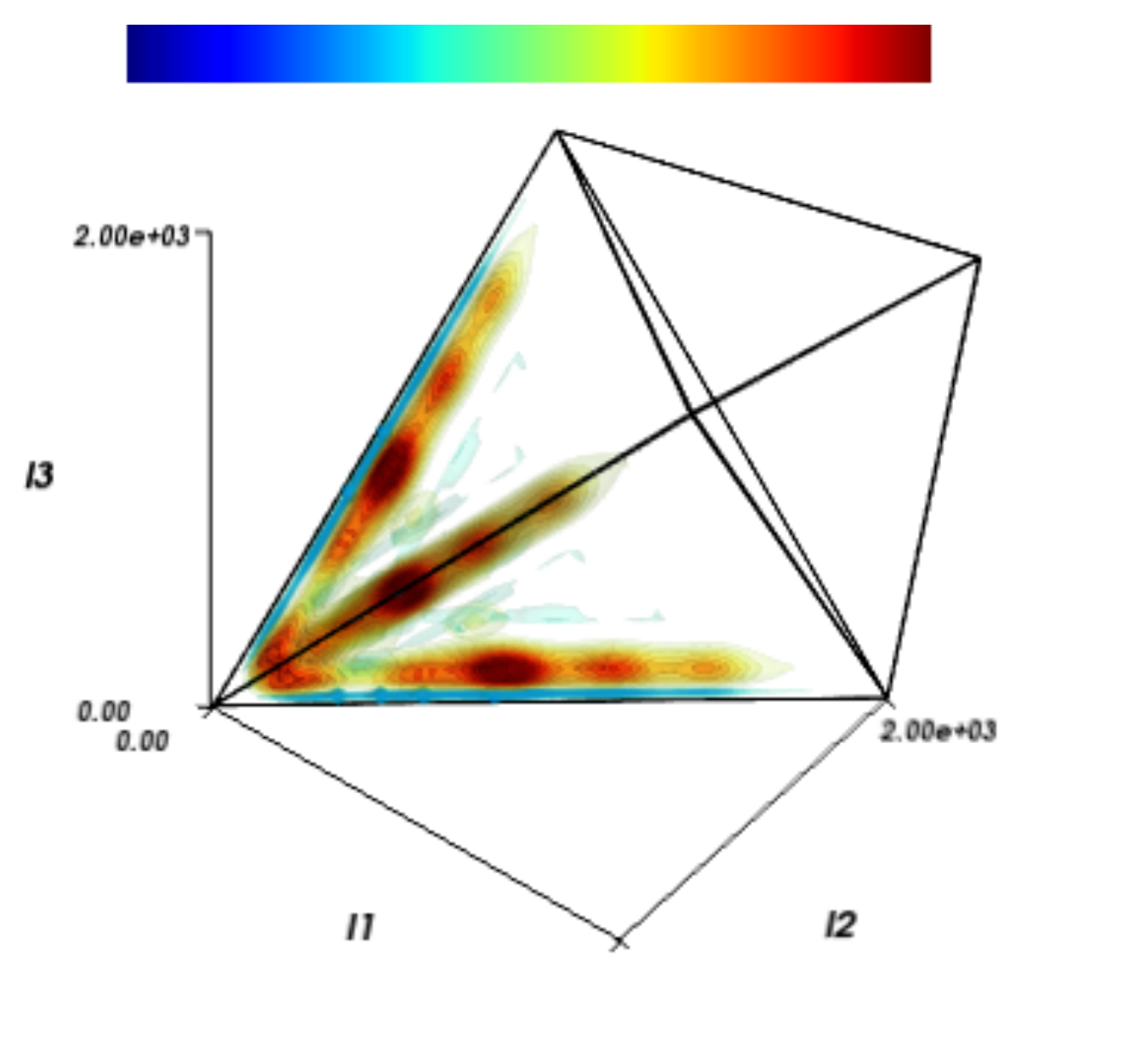}}
{\includegraphics[trim = 35mm  80mm 40mm 90mm, clip,width=0.48\linewidth]{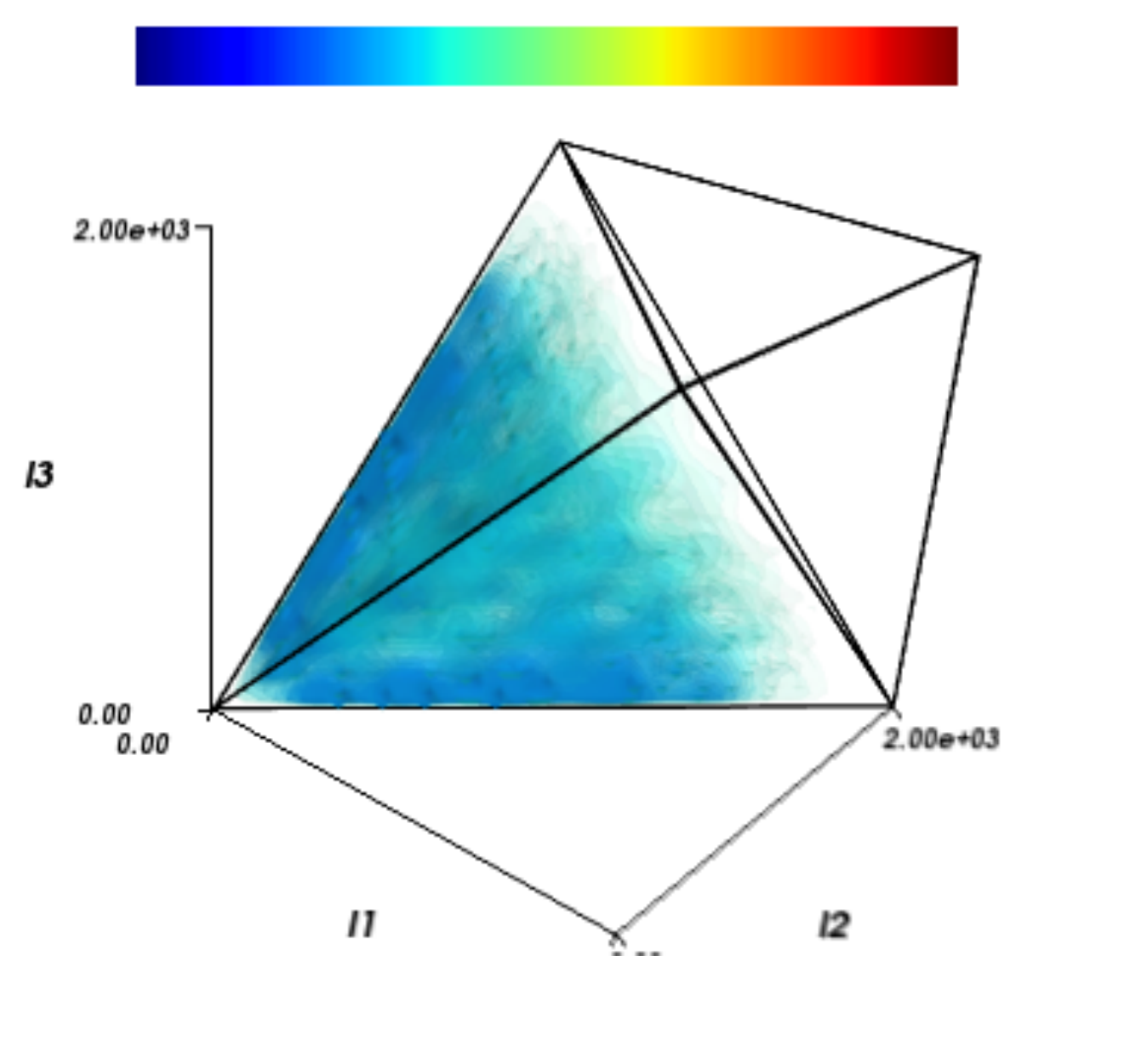}}
\caption{(left) Bispectrum of the local model bispectrum (normalised by the constant Sachs Wolfe bispectrum). The acoustic peaks are evident along with the characteristic peaking of the shape along the edges of the shape. (right) We plot the sum of the ISW, intrinsic and Doppler bispectra (normalised by the constant SW bispectrum). The shape is seen to show a mild peaking along the edges, but is largely constant (i.e. proportional to the constant SW bispectrum) and negative. The comparison to the local model emphasises the distinct signature of the cosmic string bispectrum, with the comparative absence of acoustic peaks most obvious.}
\label{fig:Bispec_Local}
\end{figure}

In order to visualise our results, we plot a series of isosurfaces of the quantity $b_{\ell_1 \ell_2 \ell_3}/b^{\rm SW}_{\ell_1 \ell_2 \ell_3}$ - normalising with respect to the constant Sachs-Wolfe bispectrum (so called as it is the CMB bispectrum resulting from a constant primordial shape $S(k_1,k_2,k_3)\equiv (k_1 k_2 k_3)^2 B(k_1,k_2,k_3)$  with the transfer function approximated by its form in the Sachs-Wolfe limit) - whose form is given by \cite{2010PhRvD..82b3502F}
\bea
b^{\rm SW}_{\ell_1 \ell_2 \ell_3} \propto \frac{1}{(2\ell_1+1)(2\ell_2+1)(2\ell_3+1)}\left[ \frac{1}{\ell_1+\ell_2+\ell_3 +3}+ \frac{1}{\ell_1+\ell_2+\ell_3 }\right]\,.
\eea
In Figure~\ref{fig:Bispec_3types} we plot the individual bispectra for the ISW, intrinsic and Doppler, noting the difference in structure between the plots which indicates a relatively low correlation, with the Doppler bispectrum of opposite sign to the other contributions. 
In Figure~\ref{fig:Bispec_Local} we plot for reference the local model bispectrum - observing the `blobby' features in the plot due to the presence of acoustic peaks and the characteristic peaking of the local shape along the edges of the `tetrapyd', i.e. at the corners of the triangles on slices of fixed scale $\ell_1+\ell_2+\ell_3$. As may be observed, the cosmic string signatures are easily distinguishable from primordial shapes such as the local model; due to the active seeding of structure, the acoustic pattern is largely washed out - even for the intrinsic and Doppler terms from recombination. The total bispectrum is also plotted in Figure~\ref{fig:Bispec_Local}. The shape is largely negative, as for the ISW only bispectrum. Due to the elimination of an $\ell^{-4}$ scaling using the constant Sachs-Wolfe bispectrum, we observe that the bispectrum is largely of a constant type shape (for $\ell\lesssim 1500$) with moderate squeezing in the edges (indicated by the deeper blue shade) - for larger $\ell$ values the bispectrum would be expected to exhibit an additional $1/\ell^2$ drop off. The shape is compared to the local bispectrum to emphasise the distinct signature - with little evidence of acoustic peaks, as expected.

In order to assess the relative magnitudes, we calculate the signal to noise \eqref{eq:signoise}, obtaining the $1\,\sigma$ bounds
\bea
{[\Delta (G\mu/10^{-6})^3]}^i &=& 0.16\,,\\
{[\Delta (G\mu/10^{-6})^3]}^d &=& 0.33\,,\\
{[\Delta (G\mu/10^{-6})^3]}^{\rm ISW} &=& 0.15\,. 
\eea
Note that the bounds are of the cube power of $G\mu$ since the bispectrum is proportional to $(G\mu)^3$.
By contrast the $1\,\sigma$ bound used by the {\it Planck} team in obtaining their signal is $[\Delta (G\mu/10^{-6})^3]_{\rm Planck}=0.2$. Therefore, the semi-analytic treatment used in this paper for the ISW signal is consistent with the detailed - and computationally highly involved - treatment used in \cite{Ade:2013xla}.

Using the correlation measure \eqref{eq:correlCMB} we find the following correlations between the three contributions
\bea
\mathcal{C}({\rm Intrinsic},{\rm Doppler})=-0.39;\quad \mathcal{C}({\rm Intrinsic},{\rm ISW})=0.03;\quad \mathcal{C}({\rm Doppler},{\rm ISW})=0.03\,,
\eea
demonstrating the low correlation between the last scattering surface contributions and the ISW bispectrum.
Combining the ISW, intrinsic and Doppler bispectra we obtain the bound 
\bea
[\Delta (G\mu/10^{-6})^3] = 0.11\,.
\eea
Thus we conclude that the power spectrum - for which the $95\%$ bound $G\mu/10^{-6} \lesssim 0.15$ is achieved \cite{Ade:2013xla} - remains a superior probe to the bispectrum when last scattering surface effects are accounted for in addition to the ISW only effect. 

One may consider whether the bispectrum may become a superior statistic for detection at higher resolution, where the GKS effect dominates the cosmic string signal. In order to estimate if this is the case the signal to noise due to the power spectrum can be compared to that of the bispectrum. More precisely once compares
\bea
\left(\frac{S}{N}\right)^2_{(2)} =\sum_\ell  \frac{(2\ell+1)}{2}\frac{C_{\ell}^2}{(C_\ell^{{\rm tot}})^2}\,,\quad \left(\frac{S}{N}\right)^2_{(3)} \approx \frac{0.117}{3}2\sqrt{\ell_{\rm max}}\sum_\ell  \frac{1}{\sqrt{\ell}} \frac{ (\ell^6 b_{\ell \ell \ell})^2}{(\ell^3 C_\ell^{{\rm tot}})^3}\,,
\eea
where we simplify the signal to noise \eqref{eq:signoise} using $h^2_{\ell_1 \ell_2 \ell_3}/\Pi_i (2 \ell_i+1)^{1/3}\approx 0.117$ for valid configurations\footnote{One may be concerned that - in this regime - the cosmic variance noise for the power spectrum (and the bispectrum) may have contributions from the trispectrum. In \cite{Hindmarsh:2009es,Regan:2009hv} it was shown that - while the contribution the trispectrum contribution may be of comparable magnitude to the power spectrum (squared) contribution - it does not dominate. Therefore, the assumption of Gaussian errors is unlikely to lead to erroneous conclusions and is adopted here for the purposes of estimation.}. Assuming for large $\ell$ - in the noiseless limit - that the power spectrum, $C_\ell^{{\rm tot}}$ is dominated by the cosmic string power spectrum, $C_\ell$, and using the fact that at high $\ell$ ($\gtrsim 3000$) the cosmic string power spectrum and bispectrum are given by the ISW only contribution (such that $\ell^3 C_\ell={\rm constant}$ and $\ell^6 b_{\ell \ell \ell}={\rm constant}$), we find that 
\bea
\left(\frac{S}{N}\right)^2_{(2)} \approx \frac{\ell_{\rm max}^2}{2}\,,\quad \left(\frac{S}{N}\right)^2_{(3)} \approx 0.16\ell_{\rm max} \frac{ (\ell^6 b_{\ell \ell \ell})^2_{\ell=3000}}{(\ell^3 C_\ell)^3_{\ell=3000}}\,,
\eea
and deduce from \eqref{eq:ISWpower} and \eqref{eq:ISWbisp} that the signal to noise of the power spectrum dominates that of the bispectrum for large values of $\ell_{\rm max}$ where the power spectrum itself may be dominated by the cosmic string contribution. This analysis has assumed a noise-free signal, while in reality the spectrum will be contaminated by point source signals, and Sunyaev-Zeldov'ich effects, which would greatly complicate the analysis. Unless such contributions correlate strongly with the cosmic string signal the conclusions presented above are not expected to change.


\section{Conclusions }\label{sec:conclusions}

In this paper we have included the last scattering surface contribution in a calculation of the CMB bispectrum induced by cosmic strings, for the first time.  To do this we have used a simplified treatment.  First, we modelled the source 3-point energy-momentum correlations using a Gaussian model of the string network, which has also been used to calculate the ISW bispectrum \cite{Hindmarsh:1993pu,Regan:2009hv}. The 3-point correlators were integrated with the Green's function for a coupled matter-radiation fluid in the tight-binding approximation.
For simplicity we used a flat sky approximation, such that the bispectrum obtained is not expected to be accurate below $\ell\lesssim200$. 

We checked our approach by comparing the angular power spectrum it produces with that of a full Boltzmann code treatment, using the Unconnected Segment Model to model the string network \cite{Pogosian:1999np}. We have verified that on scales $\ell\gtrsim 200$, the qualitative and quantitative features of the power spectrum are reproduced, giving confidence to the efficacy of our approach. 

The main goal of the paper was to assess the signal to noise of the last scattering surface bispectrum with last scattering surface effects included, compared to the ISW-only contribution. The ISW-only bispectrum produced by our approach compares well to the ISW bispectrum computed from maps of the microwave anisotropy produced by the Gott-Kaiser-Stebbins effect around a numerical simulation of Nambu-Goto strings \cite{Fraisse:2007nu,Ringeval:2010ca,Ade:2013xla}.
%
Our calculation of the bispectrum contributions uses the separable basis decomposition technique - outlined in Appendix~\ref{sec:AppA}. This allows us to very efficiently calculate the signal to noise of the various terms. 

We obtain the the $1\,\sigma$ bound $\Delta (G\mu/10^{-6})^3 =0.11$, with the ISW only bound $\\ \Delta (G\mu/10^{-6})^3_{\rm ISW} = 0.15$ consistent with that obtained with detailed and highly involved simulations performed by the {\it Planck} team \cite{Ade:2013xla}. With cosmic strings constrained using the CMB power spectrum to $G\mu/10^{-6} \lesssim 0.15$ at $95\%$ confidence, we conclude that the bispectrum remains an inferior probe to the power spectrum with which to search for cosmic strings in Planck data, even with the last scattering surface contributions included.

Finally, we examine whether the bispectrum might take over as a better statistic at higher resolution, where the sharp edges in the maps of the GKS effect are better resolved.  By comparing the signal-to-noise of an ideal cosmic string CMB power spectrum with the string bispectrum, we show that constraints from the power spectrum are always superior to those from the bispectrum.

\section*{Acknowledgements}
DR acknowledges support from the Science and Technology Facilities Council
(grant number ST/I000976/1). DR is also supported by funding from the European Research Council under the European Union's Seventh Framework Programme (FP/2007-2013) / ERC Grant Agreement No. [308082]. 
This work was undertaken on the COSMOS Shared Memory system at DAMTP, University of Cambridge operated on behalf of the STFC DiRAC HPC Facility. This equipment is funded by BIS National E-infrastructure capital grant ST/J005673/1 and STFC grants ST/H008586/1, ST/K00333X/1.

\label{acknow}

\appendix

\section{Separable Decomposition }\label{sec:AppA}
In the case of the intrinsic and Doppler terms, calculation of the CMB bispectrum using equation~\eqref{eq:CMBbisp} is an almost intractable task, in general, unless the Fourier bispectrum is separable, i.e. unless the bispectrum, $B(k_1,k_2,k_3)$ can be written as a sum of terms of the form $f(k_1)g(k_2)h(k_3)$, for arbitrary functions of one variable $f,g,h$. In addition for the case of the ISW bispectrum, while one may compute the angular bispectrum, $b_{\ell_1 \ell_2 \ell_3}$ for a series of configurations, calculation for all $\ell_i\in [ 2,2000]$ is also prohibitive. 
\par Fitting the ISW bispectrum from a grid of values to an appropriate (separable) basis allows one to infer the bispectrum at arbitrary configurations (alternatively one may utilise an interpolation routine). In \cite{2010PhRvD..82b3502F,2010PhRvD..82b3520R,2010arXiv1012.6039F} a scheme was developed to allow for an arbitrary (but sufficiently smooth) bispectrum to be decomposed into a sum of separable basis functions, satisfying our requirements. We briefly summarise the approach used here (for more details refer also to \cite{2012JCAP...12..032F,2013PhRvD..88d3512R,2015JCAP...01..013R}).

\para{Fourier Bispectrum Decompositions} The Fourier shape is defined as $S(k_1,k_2,k_3)=\\(k_1 k_2 k_3)^2 B(k_1,k_2,k_3)$, while the Dirac delta condition ensures that the bispectrum is defined on triangles for fixed values of $k_1+k_2+k_3$. Utilising appropriately defined one dimensional polynomials of order $n$, $q_n(k/k_{\rm max})$ (where $k_{\rm max}$ is the maximum wavenumber considered) - one may use Legendre modes, or, as done in this paper, those defined in \cite{2010PhRvD..82b3502F}. The three dimensional basis, $Q_n$, is given by triples of these polynomials, labelled by an index $n=\{n_1,n_2,n_3\}$ which are partially ordered. In particular, we define
\bea
Q_n(k_1,k_2,k_3)=\frac{1}{6}(q_{n_1}(k_1/k_{\rm max})q_{n_2}(k_2/k_{\rm max})q_{n_3}(k_3/k_{\rm max}) + 5\,{\rm permutations\, of\, }\{n_i\}\,.
 \eea
One may define an inner product. This measure is especially useful since it allows for the definition of a correlation function. An obvious definition is to integrate over all allowed configurations. The triangle condition sets this region to be
\bea
\mathcal{V}_k=\{k_1,k_2,k_3: \sum_i k_i \leq 2 {\rm max}(k_i), k_i\leq k_{\rm max}\}\,.
\eea
In the case of the intrinsic and Doppler bispectra we choose $k_{\rm max}=0.5\,{\rm Mpc}^{-1} $ to ensure convergence of the integration to calculate the CMB bispectrum. The inner product is given by
\bea\label{eq:Fourierinnerproduct}
\langle f g\rangle =\int_{\mathcal{V}_k}dk_1 dk_2 dk_3 f(k_1,k_2,k_3)g(k_1,k_2,k_3)\,,
\eea
with the correlation between shapes $f$ and $g$ given by ${\rm Corr}(f,g)=\langle f g\rangle/\sqrt{\langle f f \rangle \langle g g \rangle}$. Setting $\tilde{\alpha}_n=\langle S Q_n\rangle$ and $\gamma_{n m}=\langle Q_n Q_m \rangle$, one may approximate the shape, $S$, as
\bea
S(k_1,k_2,k_3)=\sum_n \alpha_n Q_n(k_1,k_2,k_3)\,,
\eea
where $\alpha_n=\sum_m \tilde{\alpha}_m \gamma_{m n}^{-1}$. Clearly the fit to the shape using the basis expansion will be improved with more modes. In this paper we use $160$ modes, for which the fit to the shapes is found to be greater than $99.5\%$ using the correlation measure defined above.

\para{CMB Bispectrum Decomposition} The Fourier decomposition allows the projection to the CMB bispectrum, \eqref{eq:CMBbisp}, for the intrinsic ($i$) and Doppler ($d$) terms to be performed in an efficient manner using
\bea\label{eq:lateexpan}
b^{(a)}_{\ell_1 \ell_2 \ell_3} = \sum_n \alpha^{(a)}_n\int dx x^2 \left(\frac{1}{6}\tilde{q}^{(a)\ell_1}_{n_1}(x) \tilde{q}^{(a)\ell_2}_{n_2}(x) \tilde{q}^{(a)\ell_3}_{n_3}(x) + 5\,{\rm permutations\, of\, }\{n_i\}\right)\,,
\eea
where $a\in \{i,d\}$ and the functions $\tilde{q}_{n}^{(a)\ell}$ are given by
\bea
\tilde{q}_{n}^{(a)\ell}(x)=\frac{2}{\pi} \int dk k^2 q_n(k/k_{\rm max}) j^{(a)}_{\ell}(k x_{\rm dec}) j_{\ell}(k x)\,,
\eea
where we recall that $j^{(i)}_\ell\equiv j_\ell$ and $j^{(d)}_\ell \equiv j'_\ell$. Writing equation~\eqref{eq:lateexpan} in the form $b^{(a)}_{\ell_1 \ell_2 \ell_3} = \sum_n \alpha_n \tilde{Q}_n$ we note that computation of the signal to noise \eqref{eq:signoise} is still a numerically challenging task since $\tilde{Q}_n$ is not an {\it a priori} separable function of the $\ell_i$ once the integration over the line of sight, $x$, is performed. While one may utilise this form, and calculate the signal to noise from each combination of the line of sight times (for each bispectrum factor), leaving the line of sight integrations to the end, a more efficient scheme was developed in \cite{2010arXiv1012.6039F,2013PhRvD..88d3512R} for carrying out this computation for general bispectra. This involves carrying out a late-time (multipole space) decomposition. As for the Fourier case, one defines an inner product, given in this case by
\bea\label{eq:inner}
\llangle f g \rrangle =\sum_{\ell_i}f_{\ell_1 \ell_2 \ell_3}g_{\ell_1 \ell_2 \ell_3} \frac{h_{\ell_1 \ell_2 \ell_3}^2}{(2\ell_1+1)^{1/3}(2\ell_2+1)^{1/3}(2\ell_3+1)^{1/3}}\,,
\eea
where $h_{\ell_1 \ell_2 \ell_3}$ is given under equation~\eqref{eq:signoise}. The denominator in the final factor in the inner product such that the weight in the inner product measure is almost constant for all configurations satisfying the triangle condition. Inspired by the signal to noise measure, \eqref{eq:signoise}, and the inner product, \eqref{eq:inner}, one defines the CMB shape, $s_{\ell_1 \ell_2 \ell_3}$, in the form
\bea\label{eq:cmbbispshape}
s_{\ell_1 \ell_2 \ell_3}=\frac{(2\ell_1+1)^{1/6}(2\ell_2+1)^{1/6}(2\ell_3+1)^{1/6}}{\sqrt{C_{\ell_1} C_{\ell_2} C_{\ell_3}}} b_{\ell_1 \ell_2 \ell_3}.
\eea
To simplify notation we set ${\bar{Q}}^{(a)}_n \equiv\tilde{Q}^{(a)}_n \Pi_{i=1}^3[(2\ell_i+1)^{1/6}/\sqrt{C_{\ell_i}}]$, such that we may write $s_{\ell_1 \ell_2 \ell_3}^{(a)} =\sum_n \alpha_n^{(a)} \bar{Q}_n(\ell_1,\ell_2,\ell_3) $.
Given the inner product one may replicate the steps in the previous section.  
One wishes to decompose the CMB shape of the intrinsic and Doppler bispectra in the form 
\bea
s_{\ell_1 \ell_2 \ell_3}^{(a)}=\sum_n \bar{\alpha}^{(a)}_n Q_n(\ell_1,\ell_2,\ell_3)\,,
\eea
such that the shape is in explicitly separable form. Performing the CMB inner product on both sides with the basis $Q_m$, results in the expression
\bea\label{eq:alphabar}
\sum_n \bar{\alpha}^{(a)}_n \gamma_{n m}^{E} = \sum_r \alpha_r^{(a)} \gamma_{r m}^{(a) L}\,,
\eea
where we use the notation $\gamma_{n m}^{E}=\llangle Q_n Q_m\rrangle$ and $\gamma_{nm}^{(a)L}=\llangle \bar{Q}^{(a)}_n Q_m\rrangle$. Therefore, the coefficients $\bar{\alpha}^{(a)}_n$ are easily found by applying the inverse $\gamma_{n m}^E$ to equation~\eqref{eq:alphabar}.

This form is useful since the signal to noise, \eqref{eq:signoise}, is given succinctly by
\bea\label{eq:signoiseexpan}
\left(\frac{S}{N}\right)^2 = \frac{f_{\rm sky}}{6} \sum_{n m}\bar{\alpha}^{(a)}_n \gamma_{n m}^E \bar{\alpha}^{(a)}_m\,.
\eea
For shapes, $s^{(a)}$ and $s^{(b)}$ with coefficients $\bar{\alpha}_n^{(a)}$ and $\bar{\alpha}_n^{(b)}$, respectively, one may also define the correlation measure, $\mathcal{C}$, 
\bea\label{eq:correlCMB}
\mathcal{C}(s^{(a)},s^{(b)}) =  \frac{\sum_{n m}\bar{\alpha}^{(a)}_n \gamma_{n m}^E \bar{\alpha}^{(b)}_m}{\sqrt{\sum_{n m}\bar{\alpha}^{(a)}_n \gamma_{n m}^E \bar{\alpha}^{(a)}_m} \sqrt{\sum_{n m}\bar{\alpha}^{(b)}_n \gamma_{n m}^E \bar{\alpha}^{(b)}_m}} \equiv \frac{\llangle s^{(a)}s^{(b)} \rrangle}{\sqrt{\llangle s^{(a)}s^{(a)} \rrangle}\sqrt{\llangle s^{(b)}s^{(b)}\rrangle} }\,.
\eea
\para{Note on the ISW Bispectrum Decomposition} 
Unlike the last scattering surface terms, the ISW bispectrum is expressed directly in multipole space using the Limber approximations of the flat-sky formula \eqref{eq:powISW}. Having calculated it on a grid of $\ell_i$ values we scale it to obtain the bispectrum shape, $s^{\rm ISW}_{\ell_1 \ell_2 \ell_3}$ using equation~\eqref{eq:cmbbispshape}. Given that the measure of the multipole space inner product is constant for almost all configurations, we instead use the more computationally efficient inner product, \eqref{eq:Fourierinnerproduct}, (where $k_i\rightarrow \ell_i$) to perform the decomposition, labelling the coefficients of the expansion to be $\bar{\alpha}_n^{\rm ISW}$, with $s^{\rm ISW}_{\ell_1 \ell_2 \ell_3}=\sum_n \bar{\alpha}^{\rm ISW}_n Q_n(\ell_1,\ell_2,\ell_3)$. The signal to noise is then obtained using equation~\eqref{eq:signoiseexpan} as for the intrinsic and Doppler terms.

\bibliography{bibli,CosmicStrings}
\end{document}